\documentclass{article}
\usepackage{geometry}
\usepackage[parfill]{parskip}
\usepackage{graphicx}
\usepackage{amssymb}
\usepackage{epstopdf}
\usepackage{natbib}
\usepackage{color}
\definecolor{dark-gray}{gray}{0.3}
\usepackage[colorlinks=true,urlcolor=dark-gray,breaklinks,citecolor=dark-gray,linkcolor=dark-gray]{hyperref}
\usepackage{breakurl}
\usepackage{subfig}
\title{Modelling ecological communities as if they were DNA}
\author{William D.\ Pearse$^{1,2,3,*}$, Andy Purvis$^{1,4}$, David B.\ Roy$^2$,and Alexandros Stamatakis$^{5,6}$}
\date{\small$^1$Department of Life Sciences, Imperial College London, Ascot, Berkshire, United Kingdom
\\\small$^2$Centre for Ecology and Hydrology, Wallingford, Oxfordshire, United Kingdom
\\\small$^3$Department of Ecology, Evolution, and Behavior, University of Minnesota, 1987 Upper Buford Circle, Saint Paul, Minnesota, 55108, USA
\\$^4$Department of Life Sciences, Natural History Museum, Cromwell Road, London SW7 5BD, UK
\\$^5$The Exelixis Lab, Scientific Computing Group, Heidelberg Institute for Theoretical Studies, Heidelberg D-69118, Germany
\\$^6$Institute for Theoretical Informatics, Karlsruhe Institute of Technology, D-76131, Karlsruhe, Germany
\\$^*$\url{wdpearse@umn.edu}
\\Date: \today}
\begin{document}
\bibliographystyle{/home/will/besjournals.bst}
\maketitle
\section{Abstract}
Ecologists are interested in understanding and predicting how
ecological communities change through time. While it might seem
natural to measure this through changes in species' abundances,
computational limitations mean transitions between community types are
often modelled instead. We present an approach inspired by DNA
substitution models that attempts to estimate historic interactions
between species, and thus estimate turnover rates in ecological
communities. Although our simulations show that the method has some
limitations, our application to butterfly community data shows the
method can detect signal in real data. Open source \texttt{C++} code
implementing the method is available at
\href{http://www.github.com/willpearse/lotto}{\url{http://www.github.com/willpearse/lotto}}.

\section{Introduction}
Many ecologists recognise broad habitat types and sub-types, grouping
communities they define as being similar in structure. A good example
is the British National Vegetation Classification system
\citep{Rodwell1991}, which hierarchically classifies plant communities
within the UK. Ecologists also recognise variability within these
categories, and the recent interest in Neutral Theory
\citep{Hubbell2001} and stochastic variation at all spatial scales
\citep[reviewed in][]{Vellend2010} suggests ecologists want to model
this variation. However, models that allow for variation within
habitat types are often over-parameterised, and summary statistics of
structure do not necessarily facilitate the prediction of future
species compositions.

One way to approach this problem has been to model a community as
proceeding through a series of states, each state representing a
particular community type with associated species compositions and
abundances. The probability of moving among these states can be
modelled using Markov chains, and thus predictions about future
ecological composition can be made \citep[reviewed in][]{Logofet2000}.
This is a natural way to model habitat types, but it cannot model
variation within states, assumes that the history of a community is
unimportant, and requires that a system reaches a final, stable
state. Moreover, such methods require \emph{a priori} definitions of
states, and as such are not solely driven by the data themselves.

Our alternative is to model species turnover as a transition matrix,
where the likelihood of an individual entering a community can be
predicted as a function of the species of the individual it
replaces. Such a model can be modified to allow the addition of
individuals without replacement, and provides both species-level
parameters of ecological interest and predictions about future species
abundances. The problem of over-parameterisation can be solved by
simplifying this matrix, analogously to procedures for simplifying DNA
or protein substitution matrices. These simplifications are applied by
allowing certain bases or amino acids (in our case species) to change
with the same rate \citep[\emph{e.g.}, the Jukes-Cantor model where
all characters evolve with the same rate;][]{Jukes1969}. However, this
method has the major drawback of requiring an accurate way of
determining the history of species interactions in a community, which,
unfortunately, is often unavailable and thus must be estimated from
species composition data.
\section{Methods}
\subsection{Overview and description of problem}
We model the fate of each individual in a community over a number of
discrete time-steps, assuming one of four events happens to each
individual in each time-step:

\begin{itemize}
\item \textbf{Reproduction}. The individual dies and is replaced by
  another of the same species (implicitly its offspring). This is
  equivalent to, and indistinguishable from, the individual surviving
  until the next time-step.
\item \textbf{Replacement}. The individual dies, and is replaced by
  another individual of a different species.
\item \textbf{Death}. The individual dies, and is not replaced by
  another individual of any species. This allows communities to
  decrease in overall abundance.
\item \textbf{Addition}. The individual enters the community, and does
  not replace any other individual. This allows communities to
  increase in overall abundance.
\end{itemize}

Although this model is conceptually straight-forward, it is difficult
to estimate the parameters involved (the rate of reproduction, loss,
death and addition); composition data do not reveal what happend
during each time-step. Taking the community in table
\ref{turnoverProblem} as an example, it is difficult to disentangle
what happened between the first and second measurements that led to
species A increasing in number and species B becoming less
abundant. Any attempt to infer what events were most likely to have
happened requires an estimate of the relative likelihoods of those
events taking place; this creates a circularity, since estimating the
likelihoods requires some knowledge of the events.
\begin{table}
\centering
\subfloat[First time-step]{
\begin{tabular}{l l}\hline
Species&Abundance\\ \hline
A&10\\
B&20\\
C&20\\ \hline
\end{tabular}}
\subfloat[Second time-step]{
\begin{tabular}{l l}\hline
Species&Abundance\\ \hline
A&20\\
B&10\\
C&20\\ \hline
\end{tabular}}
\caption{The problem of estimating species turnover. What happened in the time between the first measurement of this community (a) and the second (b)? Did ten individuals of species B become replaced by ten of species A? Did ten individuals of species B die out without leaving descendants, and ten members of species A come from outside the community? Did ten individuals of species C become replaced by ten of species A, and another ten came in from outside the community? There are many possible transitions between the two communities, and no obvious way to determine what happened without already having a model for the likelihood of possible transitions.}
\label{turnoverProblem}
\end{table}
\subsection{Description of method}
Open source \texttt{C++} code (named `\emph{lotto}') that implements
the method described below is available at
\leavevmode\href{http://www.github.com/willpearse/lotto}{\url{http://www.github.com/willpearse/lotto}}. The
program is also capable of simulating data with which to test the
method.

The method assumes that community compositions are known perfectly,
and all were repeatedly sampled at the same frequency. It starts by
generating an initial \emph{transition matrix} (table
\ref{transMatEg}), which contains the relative rates of reproduction,
replacement, death, and addition for each species. Within each row all
bar the last column must sum to one, since each individual must do
something in each time step. In the last column all the rows must sum
to one, since every time an addition takes place the individual must
be of a species.
\begin{table}
\begin{center}
\begin{tabular}{c|cccc|c}
&A&B&C&Death&Addition\\ \hline
A&reproduction&replacement&replacement&death&addition\\
B&replacement&reproduction&replacement&death&addition\\
C&replacement&replacement&reproduction&death&addition\\
\end{tabular}\\
\caption{Example \emph{transition matrix}. Each species is a
  represented by a letter (`A', `B', and `C'), and each element of the
  matrix represents a parameter of the model, as described in the
  text.}\label{transMatEg}
\end{center}
\end{table}

Using the transition matrix, \emph{lotto} must now estimate what
happened between each time-step in each community. In random order,
each individual is assigned its most likely source (a reproduction,
replacement, or addition event) given the individuals in that
community's previous time step not yet assigned to an
event. Individuals undergo death events when a community has a lesser
overall abundance than the previous time-step. This creates an
\emph{event matrix}, of the same dimensions as the transition matrix,
with counts of the number of times each possible event in the
transition matrix occurred in each community between each time-step.

We can now estimate the likelihood of the data (the event matrix)
given the model (the transition matrix). \emph{Lotto} repeats the
process of calculating the event matrix given the transition matrix,
conducting a maximum likelihood optimisation for each transition
parameter in random order using Brent's method \citep{Brent1973}. Once
each transition parameter has been optimised, the event matrix is
rescaled so that the row and column totals conform to the restrictions
above. This process of recalculating the entire transition matrix
should be iterated until the parameter estimates converge, although at
present \emph{lotto} does not assess convergence.
\section{Example with simulated data}
We present an example of this method with ten random communities, each
with ten years of data, starting with 100 individuals and having ten
individuals added at each time-step. We used \emph{lotto} to simulate
these data, and then made one attempt (with five iterations) to
estimate transition and event matrices. Tables \ref{simTrans} and
\ref{simEvent} show the true and estimated transition and event
matrices, along with the transition matrix used to initiate the
search.

It is clear that the method has flaws. Although species with higher
reproduction rates have higher estimated rates, these estimates are
inflated, the method is poor at detecting death events, and
underestimates rates of addition. This may stem from the special (but
perhaps likely) case where transition and death rates for a species
are identical, such that there may be equally likely ways of
explaining the results. Repeated searches from different, randomised,
starting transition matrices could help the program escape such local
optima. There may also be identifiability issues relating to the death
and addition rates; it is notable that the species with the lowest
real death rate (`c') has the highest estimated rate of addition.
\begin{table}
\centering
\subfloat[Real]{
\begin{tabular}{l |llllll |l}
&a&b&c&d&e&Death&Addition\\ \hline
a&{\bf0.46}&0.11&0.11&0.11&0.11&0.11&0.20\\
b&0.07&{\bf0.66}&0.07&0.07&0.07&0.07&0.20\\
c&0.02&0.02&{\bf0.88}&0.02&0.02&0.02&0.20\\
d&0.03&0.03&0.03&{\bf0.85}&0.03&0.03&0.20\\
e&0.09&0.09&0.09&0.09&{\bf0.53}&0.09&0.20\\
\end{tabular}}\\
\subfloat[Initial]{
\begin{tabular}{l |llllll |l}
&a&b&c&d&e&Death&Addition\\ \hline
a&{\bf0.20}&0.16&0.16&0.16&0.16&0.16&0.20\\
b&0.16&{\bf0.20}&0.16&0.16&0.16&0.16&0.20\\
c&0.16&0.16&{\bf0.20}&0.16&0.16&0.16&0.20\\
d&0.16&0.16&0.16&{\bf0.20}&0.16&0.16&0.20\\
e&0.16&0.16&0.16&0.16&{\bf0.20}&0.16&0.20\\
\end{tabular}}\\
\subfloat[Estimated]{
\begin{tabular}{l |llllll |l}
&a&b&c&d&e&Death&Addition\\ \hline
a&{\bf0.65}&0.01&0.12&0.11&0.02&0.01&0.23\\
b&0.03&{\bf0.84}&0.06&0.05&0.01&0.01&0.16\\
c&0.00&0.01&{\bf0.96}&0.01&0.00&0.01&0.2771\\
d&0.01&0.01&0.01&{\bf0.95}&0.01&0.01&0.01\\
e&0.01&0.06&0.04&0.052&{\bf0.82}&0.01&0.32\\
\end{tabular}}
\caption{Values of the \emph{transition matrix} used to generate the data (a), to start the search procedure (b), and given as the estimated result (c). Reproduction parameters have been highlighted. This example shows a general tendency of the program to under-estimate death, and to over-estimate the likelihood of an individual to reproduce.}\label{simTrans}
\end{table}
\begin{table}
\centering
\subfloat[Real]{
\begin{tabular}{l |llllll |l}
&a&b&c&d&e&Death&Addition\\ \hline
a&{\bf566}&121&148&150&131&136&169\\
b&125&{\bf1234}&123&111&124&122&195\\
c&88&92&{\bf2785}&79&80&63&173\\
d&108&93&92&{\bf2447}&82&85&166\\
e&126&130&137&143&{\bf786}&120&197\\
\end{tabular}}\\
\subfloat[Estimated]{
\begin{tabular}{l |llllll |l}
&a&b&c&d&e&Death&Addition\\ \hline
a&{\bf896}&12&96&93&57&8&45\\
b&84&{\bf1627}&78&24&26&0&63\\
c&11&34&{\bf3104}&27&10&1&116\\
d&22&28&33&{\bf2785}&38&1&106\\
e&34&101&31&61&{\bf1215}&0&54\\
\end{tabular}}
\caption{Values of the \emph{event matrix} used to generate the data (a), and estimated at the end of the run (b). Reproduction parameters have been emboldened. This example shows a general tendency of the program to under-estimate death, and to over-estimate the likelihood of an individual to reproduce.}\label{simEvent}
\end{table}
\section{Example with ecological data}
We used data from a well-recorded site (site 10, UK grid-reference
NT590070) from the \href{http://www.ukbms.org.uk}{UK Butterfly
  Monitoring Scheme} as a test-case of the method. We grouped the
species' abundances into five groups of butterflies \citep[the
`skippers', `whites', `hairstreaks, coppers and blues', `fritillaries
and nymphalids', and `browns' as defined in][]{Asher2001} as input for
the program to reduce the problem dimension. Grouping species
according to their taxonomy is not unreasonable given the phylogenetic
signal in species' traits and assemblage structure (WD Pearse
unpublished obs.). The results, after only one search with five
iterations, are shown in table \ref{ecolResults}.
\begin{table}
\centering
\subfloat[\emph{Transition Matrix}]{
\begin{tabular}{l |llllll |l}
&Whites&Blues&Skippers&Fritillaries&Browns&Death&Addition\\ \hline
Whites&{\bf0.38}&0.01&0.02&0.01&0.01&0.57&0.04\\
Blues&0.30&{\bf0.22}&0.04&0.01&0.07&0.35&0.03\\
Skippers&0.00&0.00&{\bf0.99}&0.00&0.00&0.00&0.01\\
Fritillaries&0.15&0.01&0.08&{\bf0.15}&0.15&0.45&0.68\\
Browns&0.01&0.00&0.01&0.00&{\bf0.95}&0.01&0.24\\
\end{tabular}}\\
\subfloat[\emph{Event Matrix}]{
\begin{tabular}{l |llllll |l}
&Whites&Blues&Skippers&Fritillaries&Browns&Death&Addition\\ \hline
Whites&{\bf16016}&590&76&64&1052&3556&2408\\
Blues&497&{\bf2208}&237&153&481&119&443\\
Skippers&380&61&{\bf863}&6&188&107&242\\
Fritillaries&168&30&71&{\bf358}&73&40&78\\
Browns&2039&386&83&98&{\bf6495}&930&1759\\
\end{tabular}}\\
\caption{Values of the \emph{transition} and \emph{event} matrices when the method was applied to real data. For ease of reading, the names of two of the groups are abbreviated to `blues' and `fritillaries' from `hairstreaks, coppers and blues' and `fritillaries and nymphalids', respectively. See text for further discussion, but note that these parameters do not resemble those of the simulated data, in particular having large `death' parameter estimates.}\label{ecolResults}
\end{table}

Given the simulated results it is not appropriate to draw strong
conclusions about these empirical results. However, a general point
can be made: these transition matrices are more variable than those
estimated from the simulated data, suggesting the method may be
picking up some signal (biological process) in the data. The low
replacement parameters suggest that few of the clades seem to be
interacting with other clades, with the exception of the blues
replacing the whites.
\section{Further work}
The simulated example shows that the method is imperfect, and makes it
clear that more work is needed. The butterfly data may still represent
a good test case for the method once it has been refined, since few
butterflies have overlapping generations and, in its current form, the
method assumes there is no generational overlap.

Although the order in which individual events are assigned to the
event matrix is randomised in each iteration of the program, the
assignment of events according to whichever is the most likely is
deterministic. This is another possible explanation for the inaccurate
estimation of rates in the simulated data, and so more explicitly
incorporating stochastic processes is an important next
step. Preliminary experiments using MCMC (Markov Chain Monte Carlo;
not shown) to both integrate over uncertainty and estimate confidence
in model parameters suggest that this is a promising avenue of future
work.

Currently, the method cannot be used with more than a handful of
species or it takes too long to parameterise the event matrix, a step
that needs further computation optimisation. However, allowing the
user to specify a number of candidate restricted transition matrices
that share parameters across species would not just allow more species
to be analysed, but also allow explicit ecological hypotheses to be
tested.
\section{Acknowledgements}
WDP was funded by the Heidelberg Institute for Theoretical Studies and
a NERC CASE PhD studentship, and is grateful for advice and comments
from A.\ Aberer, C.\ Goll, F.\ Izuierdo-Carrasco, K.\ Kobert, K.\
Luckett, P.\ Pavlidis, and S.\ Pissis, and J.\ Zhang, in
Heidelberg. B.\ Waring gave useful feedback on the manuscript. We are
also greatful to all those involved in the collection and collation of
the UKBMS data.

\end{document}